\documentclass[proof]{WileyASNA-v1}

\articletype{Article Type}%

\usepackage{epsfig}
\usepackage{amsmath}

\usepackage{graphicx}

\received{05 December 2018}
\revised{10 December 2018}
\accepted{xxx}

\begin{document}

\title{Conformal Gravity and the Radial Acceleration Relation}

\author[1]{James G. O'Brien*}
\authormark{James G. O'Brien et. al.}
\author[2]{Thomas L. Chiarelli}
\author[3]{Mark A. Falcone}
\author[3]{Muhannad H. AlQurashi}

\address[1]{\orgdiv{Department of Mathematics, Physics and Computer Science}, \orgname{Springfield College}, \orgaddress{\state{263 Alden St., Springfield, MA}, \country{USA}}}
\address[2]{\orgdiv{Department of Electrical and Mechanical Engineering}, \orgname{Wentworth Institute of Technology}, \orgaddress{\state{550 Huntington Ave., Boston, MA}, \country{USA}}}
\address[3]{\orgdiv{Department of  Mechanical Engineering}, \orgname{Wentworth Institute of Technology}, \orgaddress{\state{550 Huntington Ave., Boston, MA}, \country{USA}}}

\corres{*Schoo-Bemis Science Center, 263 Alden St., Springfield, MA, USA.  \email{jobrien7@springfield.edu}}

\abstract{During the 2016 International Workshop on Astronomy and Relativistic Astrophysics (IWARA), the question was raised as to if conformal gravity could explain the timely result of McGaugh et. al. 2016 which showed a universal nature found in the centripetal accelerations of spiral galaxies.  At the time of the conference, the McGaugh result was only published for two weeks.  Since then, the result has become known as the Radial Acceleration Relation (RAR) and has been considered tantamount to a natural law.  In this work, we summarize how conformal gravity can explain the Radial Acceleration Rule in a fashion consistent with the findings of the original authors without the need for dark matter.}


\jnlcitation{\cname{%
\author{O'Brien,~J.~G. et.al.}} (\cyear{2019}), 
\ctitle{Conformal Gravity and the Radial Acceleration Relation}, \cjournal{Astronomische Nachrichten}, \cvol{2019;}.}

\maketitle

\section{Introduction}
The missing matter problem in large scale physics is one that has been persisting through the turn of the century.  With radio, x-ray and optical telescope strengths increasing, data driven observations of the rotation curves of spiral galaxies have shown that the missing matter problem is not due to inadequate technology.  The most widely accepted explanation of the discrepancy in rotation velocities of galaxies has been attributed to Cold Dark Matter and can be modeled using formalisms such as \cite{nfw}.  However, due to the lack of direct observational detection, more authors than ever have been exploring the possibility that the current Einstein Gravity may be in need of a modification or replacement to solve the missing matter problem.  Although there are now many potential alternative gravitational theories in the literature, we will focus our efforts here to Conformal Gravity (CG).  A summary of the history and context of CG will be discussed in section \ref{cghis}.\\
Any viable theory (standard gravity with dark matter or an alternative) must be able to address observational phenomena.  Aside from the flattening rotation curves which has been ascribed to dark matter, there are other empirical relations that are of interest to the field. For example, the Tully-Fisher (TF) relation shows that for almost all observed rotation curves, there exists the relationship of $M_{disk} \propto v_{OBS}^4 $ in the outer regions of spiral galaxies.  Recently, a new empirical phenomena was discovered by \cite{mcgaughprl}  which shows a strong correlation between the predicted centripetal accelerations due to luminous Newtonian matter alone ($g_{bar}$) and the observed centripetal accelerations ($g_{OBS}$).  Here, the centripetal accelerations are the usual {$g_{OBS}=\frac{v^2_{OBS}}{R}$} for a given observed velocity data point $v_{OBS}$ at a distance $R$ from the center of the galaxy.  Taking these values with a corresponding expectation value of the centripetal accelerations {$g_{bar}=\frac{v_{bar}^2}{r}$}, across every point in a rotation curve (and then across a survey of rotation curves), one can create a plot of the points ($g_{bar},g_{OBS}$).  When viewed at an appropriate scale (see \cite{onelaw} ), the  ($g_{bar},g_{OBS}$) plot highlights the correlation between observation and prediction of the luminous matter, and has become known as the Radial Acceleration Relation (RAR).  Further, a fitting function is used,

\begin{equation}
g_{OBS}=\mathcal{F}(g_{bar})=\frac{g_{bar}}{1-e^{-\sqrt{g_{bar}/g_{\dagger}}}},
\label{E1}
\end{equation}
where  {$g_{\dagger}=1.2*10^{-10}ms^{-2}$} is the best fit free parameter found by McGaugh et al. from the chosen data set.  The correlation described in Equation (\ref{E1}) is independent of any given theory, and should also be minimally\footnote{We note here that the value McGaugh et al. obtained for $g_{\dagger}$ is directly related to the data set chosen, as well as any assumptions that may have gone into the fitting.  Specific details will follow in section \ref{rplotscg}.} coupled to any data set.  McGaugh et. al. used the Spitzer Photometry and Accurate Rotation Curves (SPARC) database (\cite{sparc}) spanning a diverse population of 153 galaxies with a total of 2693 data points.  One conclusion of \cite{mcgaughprl} is that eq. (\ref{E1}) can be used to constrain the dark matter required in rotation curve physics since there is not a one to one correlation between $g_{OBS}$ and $ g_{bar} $, namely that {$g_{dm}=g_{OBS}-g_{bar}$} .  Another conclusion discussed in  \cite{mcgaughprl} is that the correlation seen in the plots of ($g_{bar},g_{OBS}$) may point towards new physics that could explain the phenomena and the apparent emergence of a fitting relation eq. (\ref{E1}).  It is this latter conclusion that will be explored in this review, namely how conformal gravity could describe the new physics required in the RAR.

\section{Review of Conformal Gravity Rotation Curve Fitting}
\label{cghis}
Conformal Gravity, also known originally as Weyl Gravity, has enjoyed success in recent years in its ability to describe and predict rotation curve physics without the need for dark matter.  The total number of galaxies fit by CG has now been expanded to over 230 with a diverse range of morphologies.  It is no surprise that CG yields a different prediction for large scale physics than standard gravity since CG is derived from the variation of the action of the conformal Weyl Tensor $C_{\lambda\mu\nu\kappa}$ , 
\begin{equation}
 \mathcal{I}=-\alpha_g\int d^4x (-g)^{1/2}C_{\lambda\mu\nu\kappa} C^{\lambda\mu\nu\kappa} ,\\
 \end{equation}
 as opposed to the standard Einstein-Hilbert Action.  Since there are terms shared by the Weyl Tensor and the Reimmann Tensor, many properties of standard gravity will emerge in the equations of motion along with some new dynamics.  The resulting field equations can be solved for the CG equivalent of a Schwarzchild Metric and then applied to the morphology of a galaxy.  The resulting formula in CG where the predicted velocity is a function of the distance from the center of the galaxy can be modeled as:

\begin{equation}
v_{CG}(R) = \sqrt{{\frac{M}{M_\odot}v_{gr}^2(R)+\frac{M}{M_\odot}F(R)+\frac{\gamma_0c^2R}{2}-\kappa c^2R^2}},
\label{total}
\end{equation}
such that 
\begin{equation}
F(R)=\frac{\gamma^*c^2R^2}{2R_0}I_1\left(\frac{R}{2R_0}\right)K_1\left(\frac{R}{2R_0}\right).\\
\end{equation}
In eq. (\ref{total}), $v_{gr}$ is the standard formula of \cite{freeman}.The constants in eq. (\ref{total}) are  ${\gamma^*=5.42e^{-41} {cm}^{-1}}$, $\gamma_0=3.06e^{-30} {cm}^{-1}$ and 
$\kappa=9.54e^{-54}~{cm}^{-2}$, $M$ is the mass of the galaxy (measured in solar masses $M_\odot$) and $R_0$ is the disk scale length of the galaxy.  The complete derivation of eq. (\ref{total}), as well as the determination of the constants (independent of any particular galactic survey) can be found in \cite{fitting}. Other features of the fitting process, such as inclusion of bulges when applicable and galactic gas are assumed, and details can be found in \cite{impact}.

\section{The RAR Plots in Conformal Gravity}
\label{rplotscg}
\begin{figure}
 \centering
 \epsfig{file=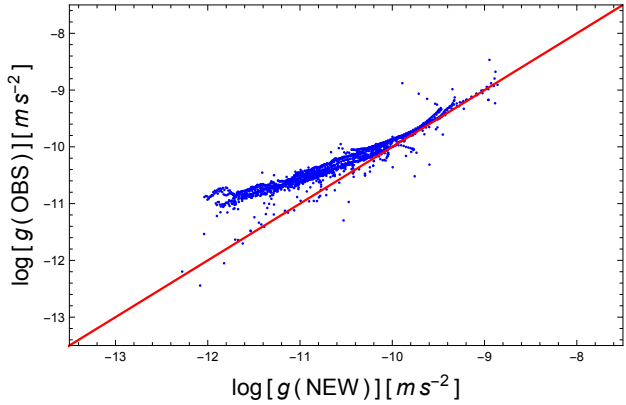,width=3.2in,height=2.05in}
  \caption{Plot of the 1614  $(g_{NEW},g_{OBS})$ points for the 40 galaxy sample. The solid line is the line of unity.}
   \label{totplot}
\end{figure}

Since the RAR is independent of any galactic survey, then without loss of generality, we present a selection of 40 galaxies\footnote{of the 40 galaxies chosen, 15 galaxies are high surface brightness (HSB) consisting of 909 points and 25 galaxies are low surface brightness (LSB) consisting of 705 points to create a fairly robust and balanced overall set.}, consisting of 1614 data points.  These particular galaxies were chosen to span the full range of galactic rotation curves from large bulged spirals to small gas dominated dwarfs.  The chosen set includes ultra high resolution data from the THINGS and LITTLE THINGS galaxies which were not included in the SPARC collection.  Figure \ref{totplot} shows the 1614 data points plotted as ($g_{NEW},g_{OBS}$) as in \cite{mcgaughprl}.  The authors here choose to use the notation of $g_{NEW}=\frac{v^2_{gr}}{R}=g_{bar}$ instead of $g_{bar}$ as in \cite{mcgaughprl} for a couple of significant reasons.  First, the data presented here does have some overlap with SPARC, but since different galaxies were used, we wanted to distinguish between the two sets.  Second, in constructing the fits shown in this work, the authors here did not restrict mass to light ratios as in \cite{mcgaughprl}.  Instead, the Nasa Extragalactic Database (NED) average distances (using cepheids when available) were used to generate the mass fitting, and all masses for the 40 galaxies were found to be of order unity.  This is a contrast with \cite{mcgaughprl} who chose to restrict the mass to light ratios, $\frac{M_\odot}{L_\odot}=.7$ for HSB galaxies and $\frac{M_\odot}{L_\odot}= .5$ for LSB galaxies.\\
 
\begin{figure}
 \centering
 \epsfig{file=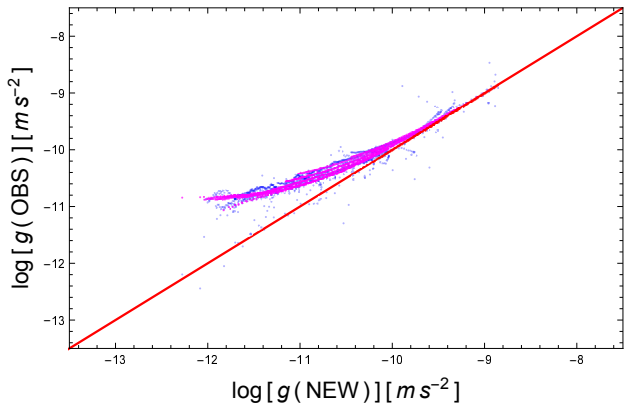,width=3.2in,height=2.05in}
  \caption{Plot of of the same 1614 points shown in Fig.\ref{totplot} with the CG prediction overlaid.}
   \label{cgcover}
\end{figure}

\begin{figure}
 \centering
 \epsfig{file=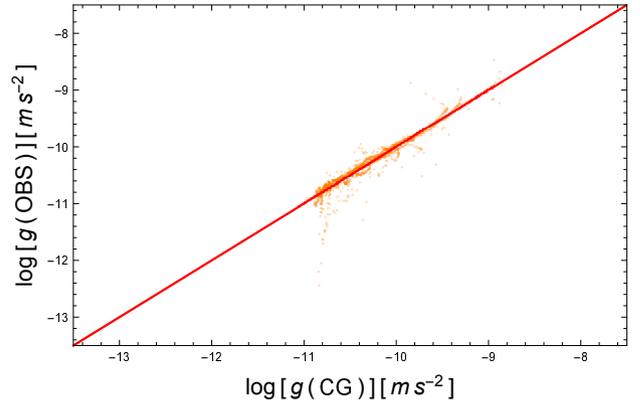,width=3.2in,height=2.05in}
  \caption{Plot of the 1614 $(g_{CG},g_{OBS})$ points for the 40 galaxy sample.  The solid line is the line of unity. }
   \label{cgplots}
\end{figure}

Figure \ref{totplot} shows the correlation between the luminous matter prediction and observation, where standard gravity would still require dark matter to overcome the discrepancy, {$g_{dm}=g_{OBS}-g_{NEW}$}.  However, we wish to test the conclusion of McGaugh et. al. that perhaps new physics could explain the apparent correlation using conformal gravity.  Applying eq. (\ref{total}) to rotation curves in general has proven successful for the 230 plus galaxies fit by conformal gravity.  The linear and quadratic terms that appear in the velocity prediction are unique to CG and depend only on constants and the relative mass of the system. There are no other free parameters that can be changed or modified when modeling.  Using eq. (\ref{total}) one can generate  {$g_{CG}=\frac{v_{cg}^2}{R}$}.  We show the results across the 1614 data points in two ways.  Figure \ref{cgcover} shows the same data in Fig. \ref{totplot} in blue, but with the conformal gravity prediction $g_{CG}$ overlaid in purple.  This plot shows that the CG prediction almost completely covers the data.  Figure \ref{cgplots} shows an alternative method for viewing the CG prediction. In Fig. \ref{cgplots}, we replace the x axis of the standard plot with $g_{CG}$ instead of $g_{NEW}$ to plot the points $(g_{CG},g_{OBS})$.  This plot is illustrative since CG makes different predictions than standard gravity. Hence, the RAR plot for CG should be the one constructed in Fig. \ref{cgplots}.


\section{Discussion and Extensions}
\label{disc}
\begin{figure*}[tbh]
\centering
\epsfig{file=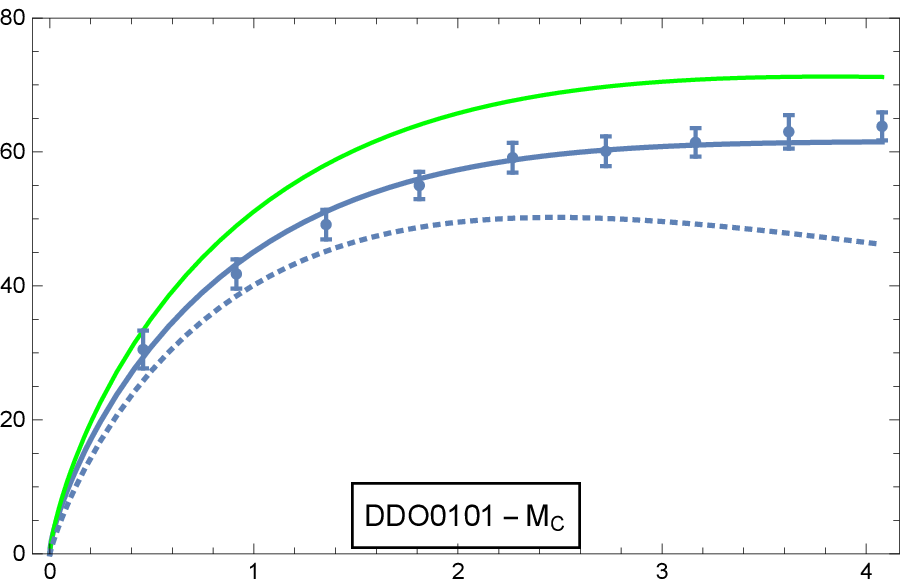,width=1.5825in,height=1.2in}~~~
\epsfig{file=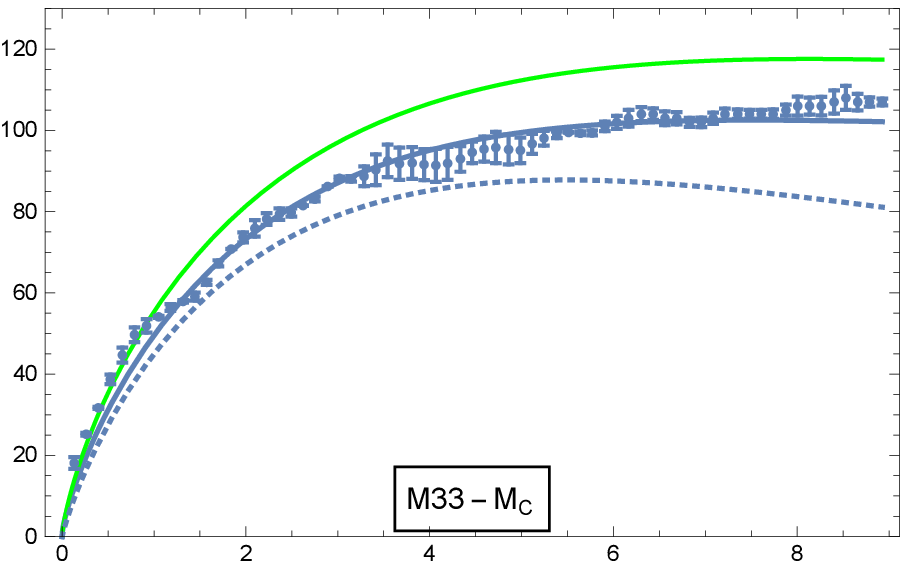,width=1.5825in,height=1.2in}~~~
\epsfig{file=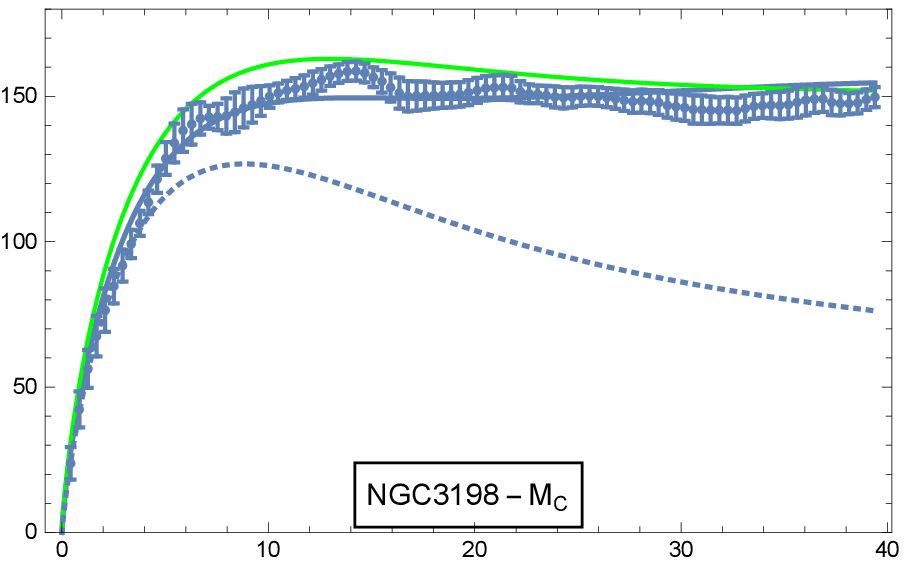,width=1.5825in,height=1.2in}~~~
\epsfig{file=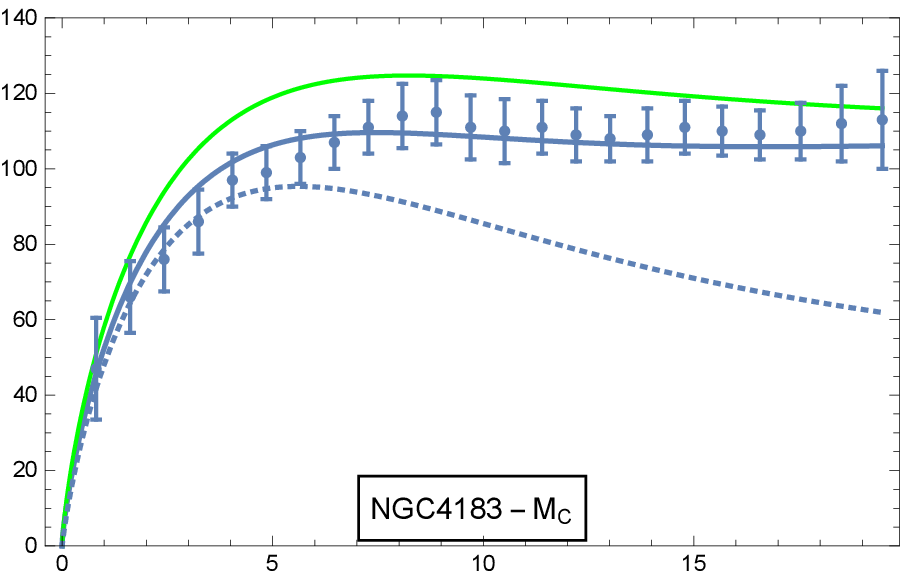,width=1.5825in,height=1.2in}\\
\smallskip
\epsfig{file=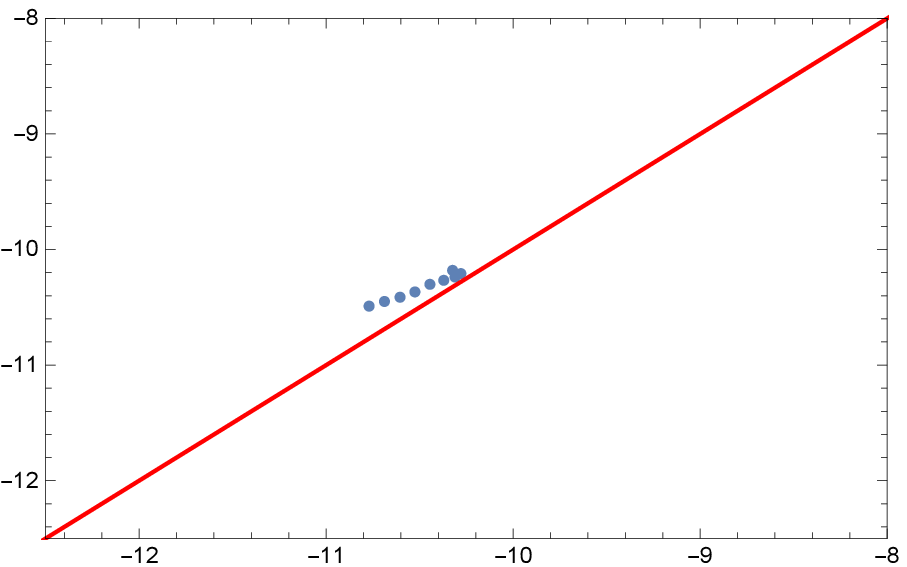,width=1.5825in,height=1.2in}~~~
\epsfig{file=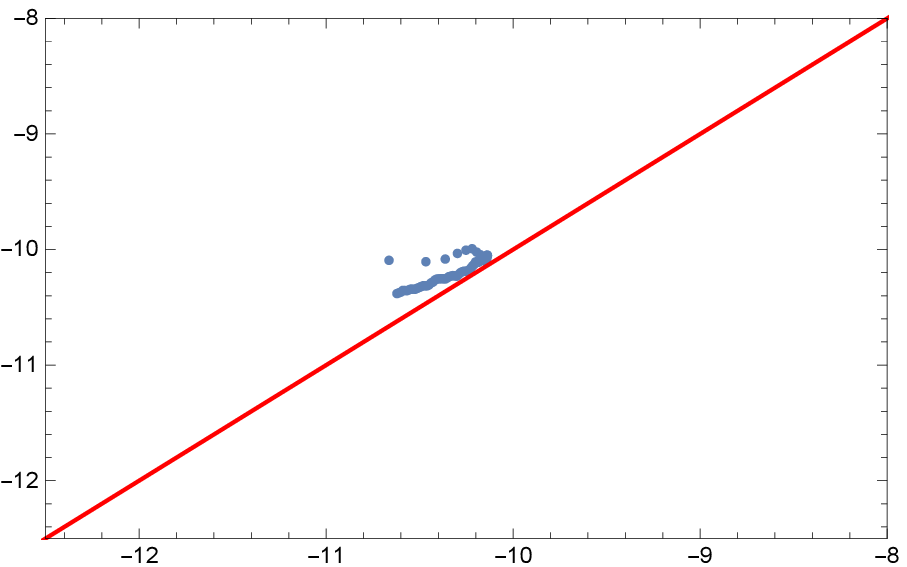,width=1.5825in,height=1.2in}~~~
\epsfig{file=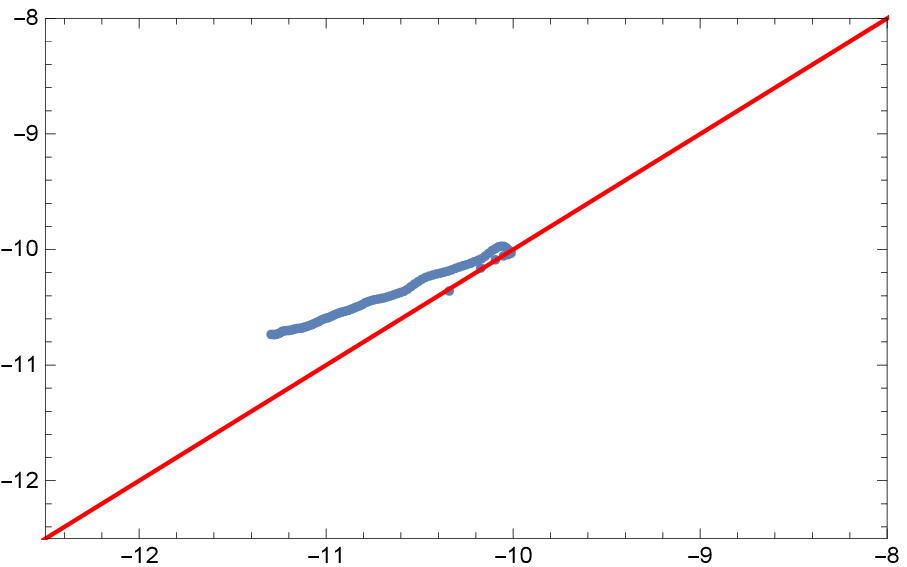,width=1.5825in,height=1.2in}~~~
\epsfig{file=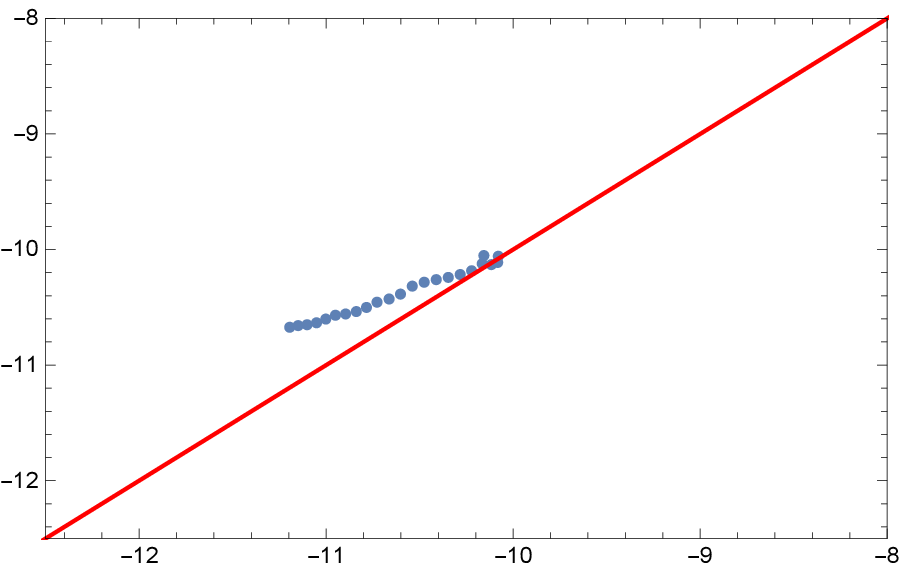,width=1.5825in,height=1.2in}\\
\smallskip
\epsfig{file=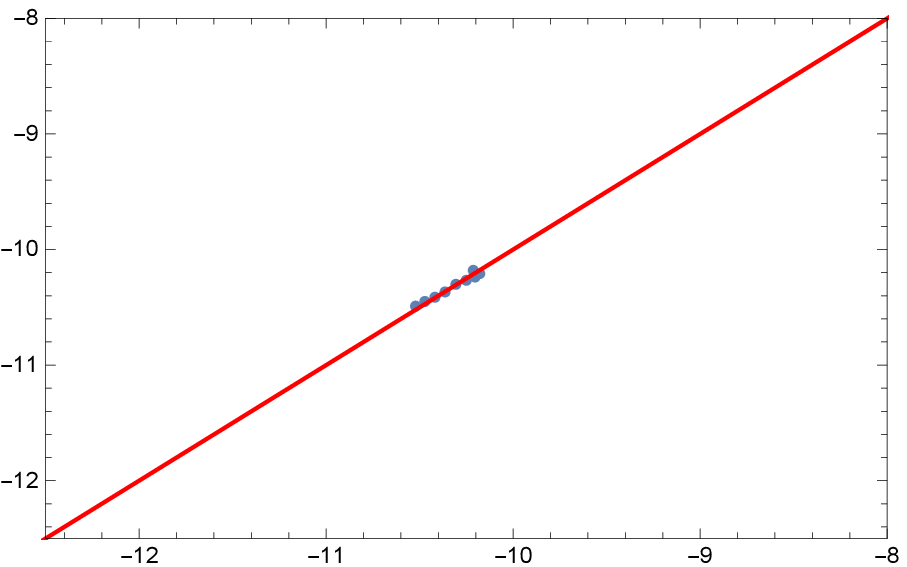,width=1.5825in,height=1.2in}~~~
\epsfig{file=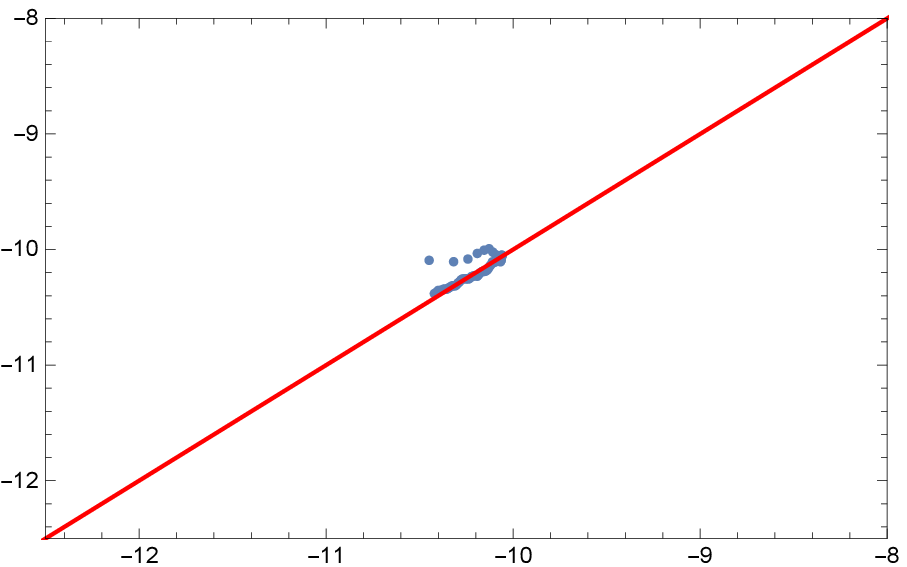,width=1.5825in,height=1.2in}~~~
\epsfig{file=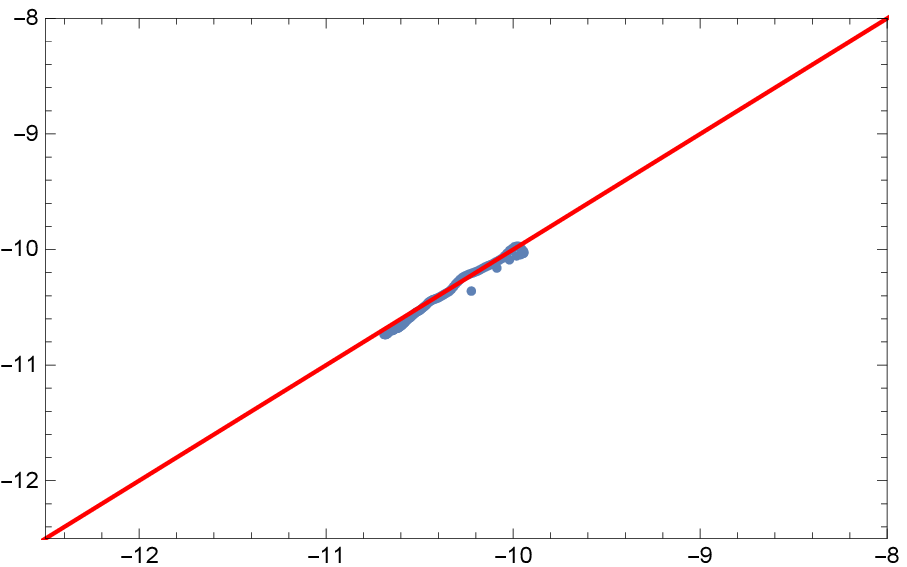,width=1.5825in,height=1.2in}~~~
\epsfig{file=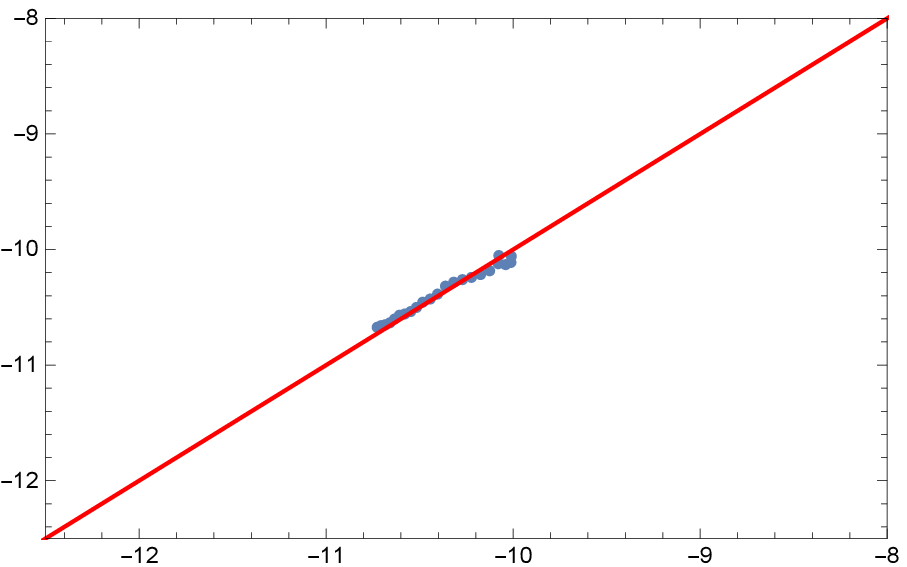,width=1.5825in,height=1.2in}\\
\medskip
\epsfig{file=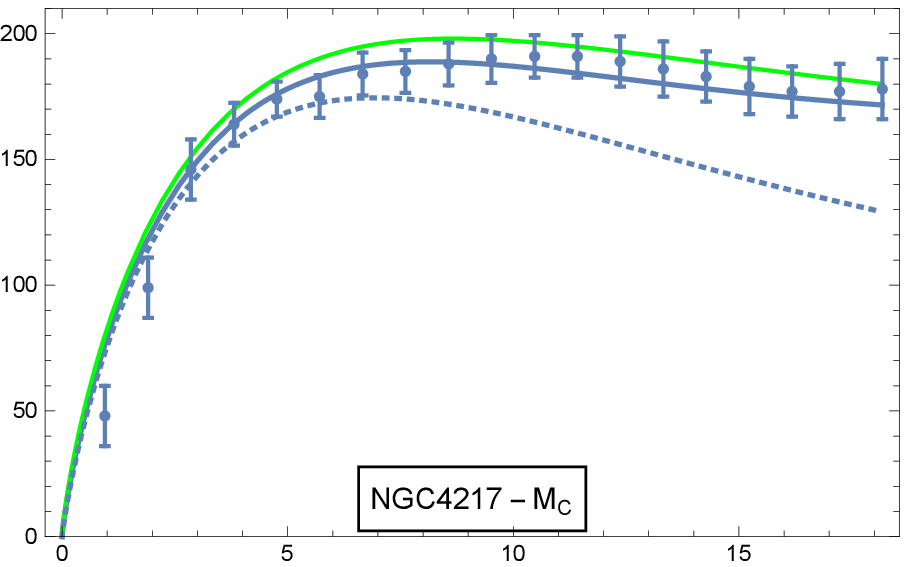,width=1.5825in,height=1.2in}~~~
\epsfig{file=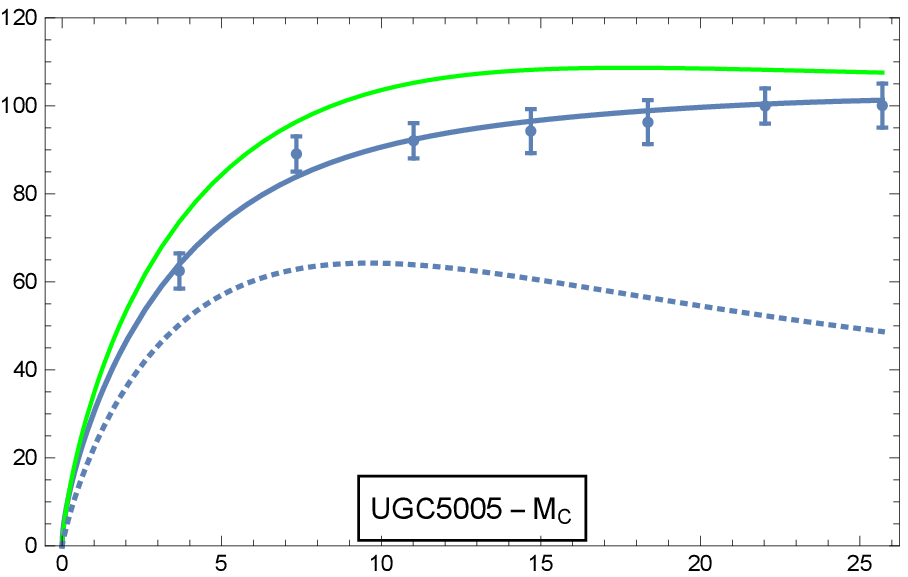,width=1.5825in,height=1.2in}~~~
\epsfig{file=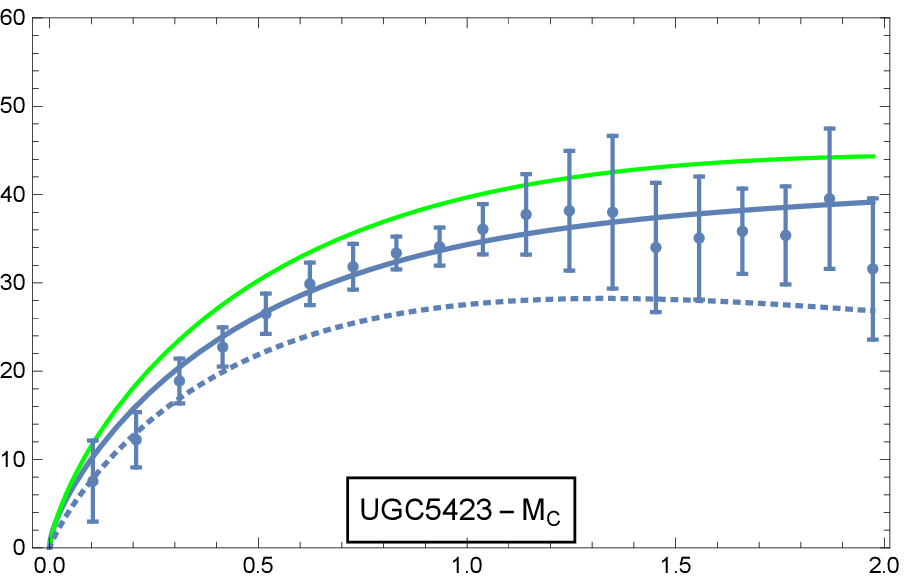,width=1.5825in,height=1.2in}~~~
\epsfig{file=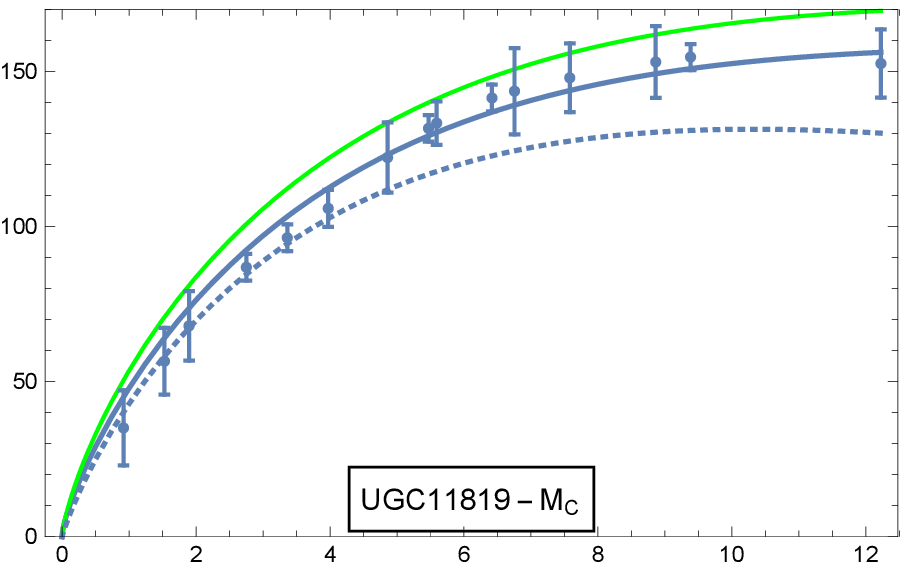,width=1.5825in,height=1.2in}\\
\smallskip
\epsfig{file=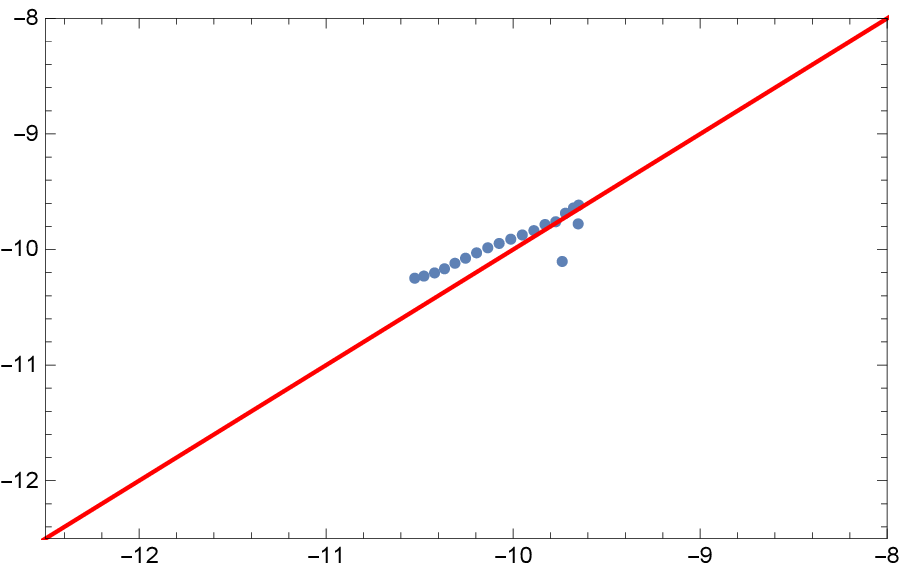,width=1.5825in,height=1.2in}~~~
\epsfig{file=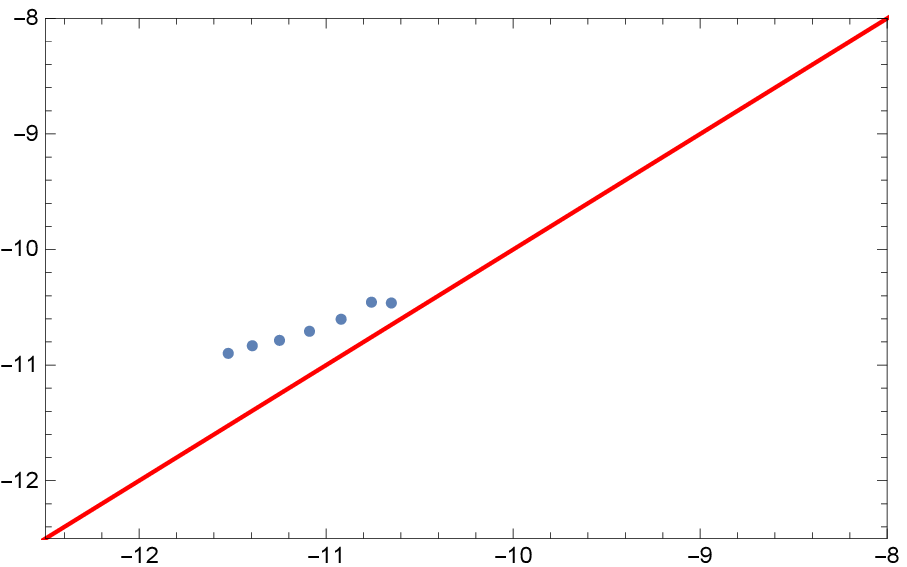,width=1.5825in,height=1.2in}~~~
\epsfig{file=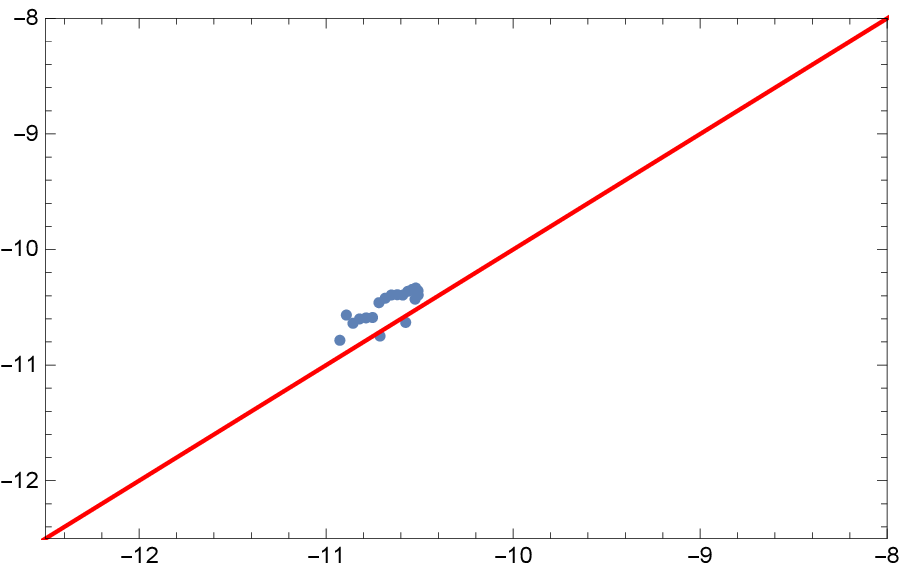,width=1.5825in,height=1.2in}~~~
\epsfig{file=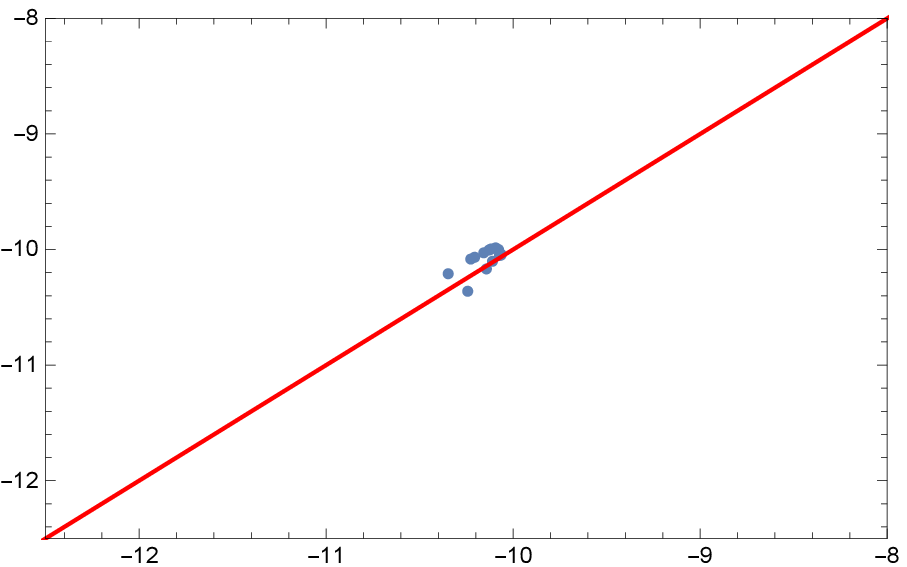,width=1.5825in,height=1.2in}\\
\smallskip
\epsfig{file=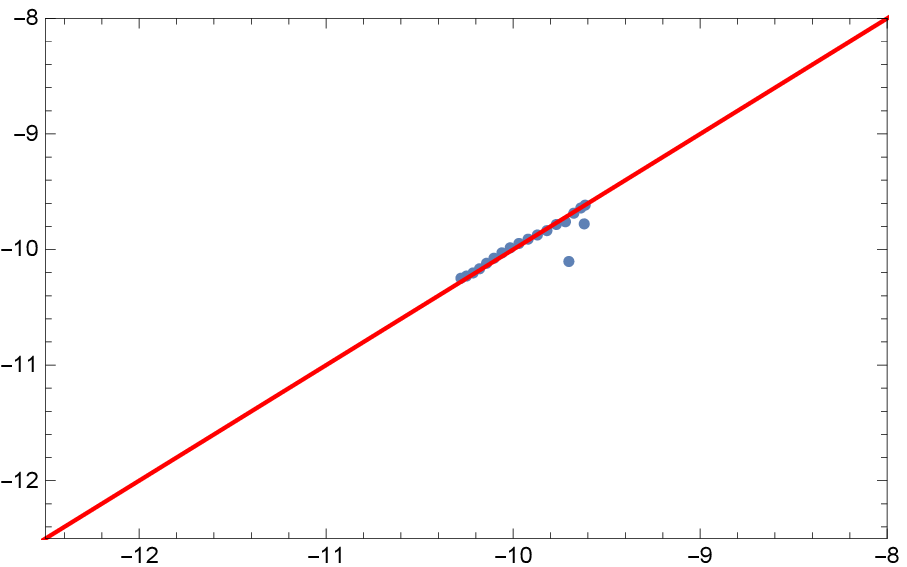,width=1.5825in,height=1.2in}~~~
\epsfig{file=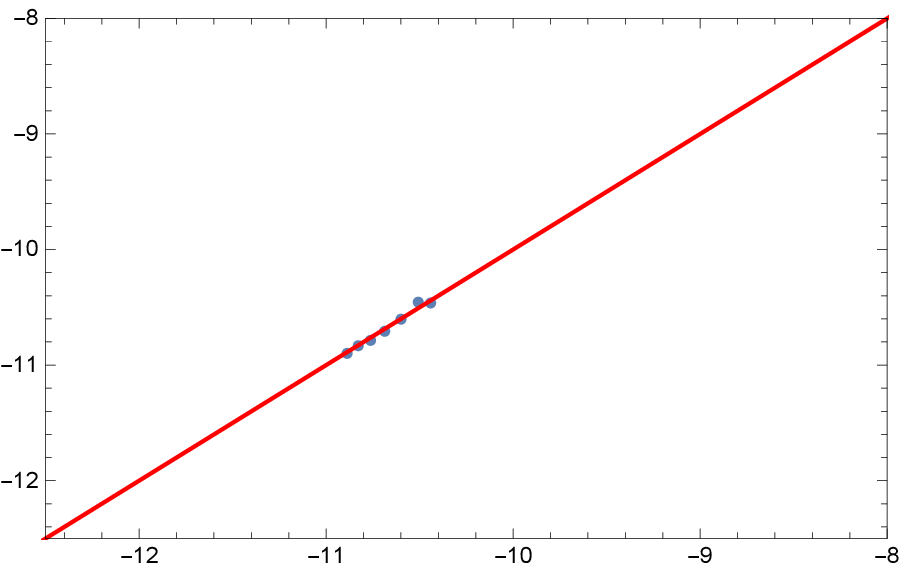,width=1.5825in,height=1.2in}~~~
\epsfig{file=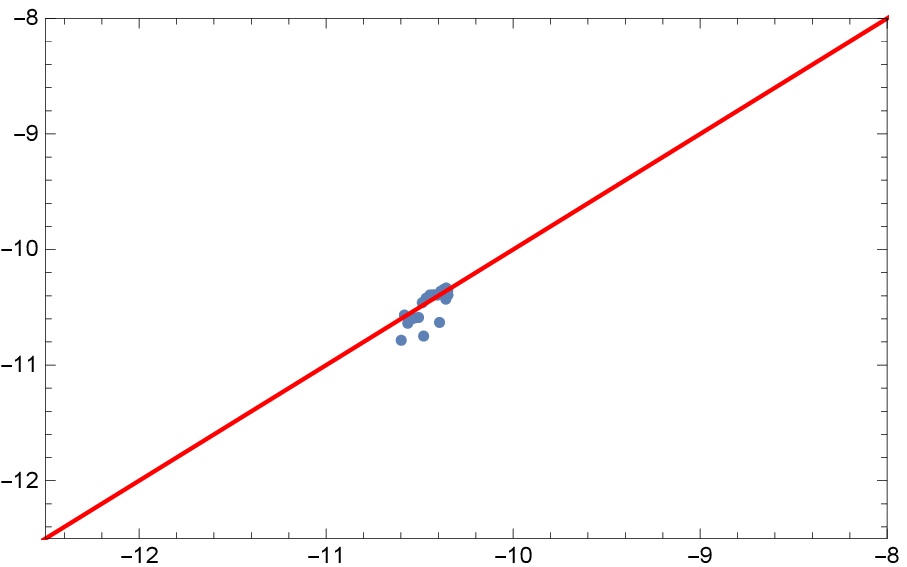,width=1.5825in,height=1.2in}~~~
\epsfig{file=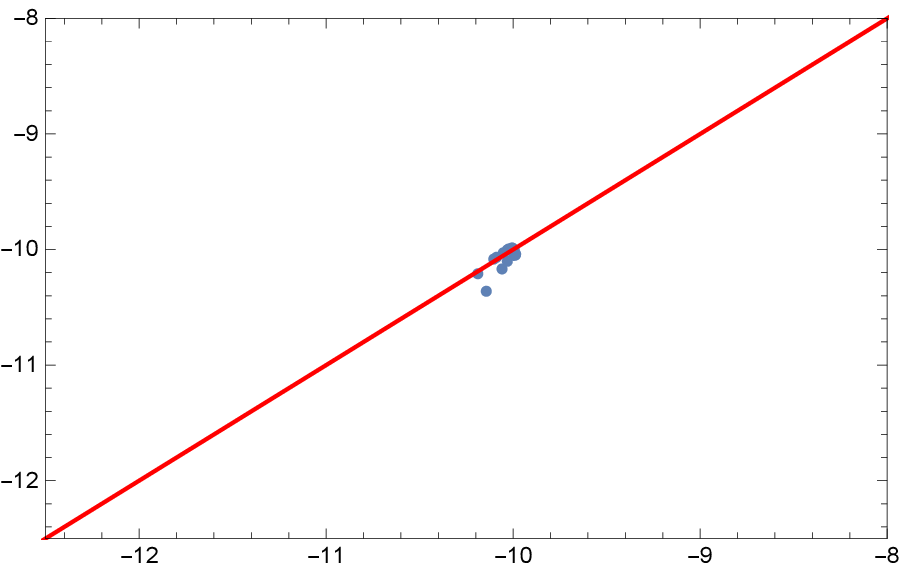,width=1.5825in,height=1.2in}\\
\smallskip
\caption{Fitting to the rotational velocities (in ${\rm km}~{\rm sec}^{-1}$) of the selected eight galaxy sample with their reported errors as plotted as a function of radial distance (in ${\rm kpc}$) is given in rows one and four. The dashed line shows the luminous Newtonian contribution, while the full curve is the conformal gravity fit.  Rows two and five show the respective contributions of the eight galaxies to Fig. \ref{totplot}, and rows three and six show their respective contributions to Fig. \ref{cgplots}.}
\label{rotcurves}
\end{figure*}

The plots shown in Fig.\ref{cgplots} highlight how CG can capture the essence of the Radial Acceleration Relation without the need for dark matter.  Further, since CG both covers the data in Fig. \ref{totplot} and makes a prediction in Fig.\ref{cgplots} that shows the correlation to be a one to one relation between luminous matter alone $g_{CG}$ and observation $g_{OBS}$, there is no need for an equivalent to eq. (\ref{E1}) in CG.  Instead, with the absence of dark matter, CG predicts that {$g_{dm}=g_{OBS}-g_{CG}=0$} .  This can be seen in Fig. \ref{cgplots} where $97\%$ of the 1614 points lie along the line of unity.  This fact is not only supportive of the claim that CG can describe the RAR, but without a fitting function as in eq. (\ref{E1}), one does not have to worry about a specific value of a fitting parameter $g_\dagger$.  This is important since the phenomena should be independent of a chosen data set. It was shown in \cite{jgoprb} that depending on the masses and distances used, the value for $g_\dagger$ can vary by up to a factor of four, thus adding to the universal power of conformal gravity.  To illustrate how the individual rotation curve of each galaxy is used in constructing the RAR plots, we present in Figure \ref{rotcurves}, the fits of eight of the chosen galaxies.  The rotation curve of each galaxy in Figure \ref{rotcurves} is followed (below) by a plot showing the respective data points contributions to the $(g_{NEW},g_{OBS})$ as well as its contribution to the plot of $(g_{CG},g_{OBS})$.  Since \cite{mcgaughprl} used eq. (\ref{E1}) to show the universal trend in the overall data, we present in each fit how eq. (\ref{E1}) applies to each individual galaxy for this data set (using the NED distances and derived masses from fitting as described above). \\
 \begin{figure}
  \centering
 \epsfig{file=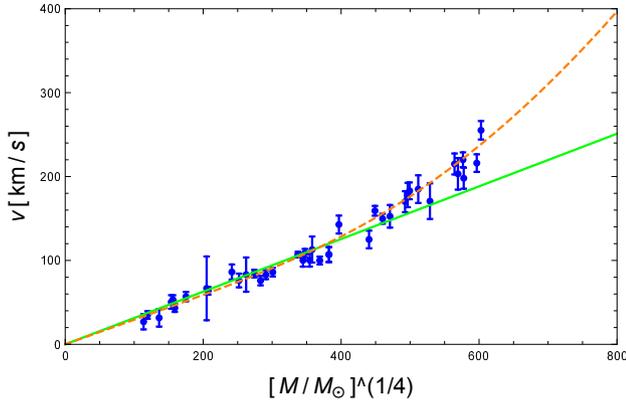,width=3.2in,height=2.05in}
  \caption{The Baryonic Tully Fisher (BTF)  plot for the last data points of the 40 galaxy sample, using the CG predicted mass. Data points are displayed with their natural error. The solid green line shows $v=M^{\frac{1}{4}}$ while the orange dashed plot shows the exact CG prediction.}
   \label{tfplot}
\end{figure}
Since the RAR is an empirical finding, a theory which explains the RAR should also be able to explain other empirical trends in the data. To this end, we present in Fig. \ref{tfplot} a plot of the Tully Fisher relationship\footnote{It should be noted that we are using the Baryonic Tully Fisher relationship (BTF), which uses the full mass of the galaxy (stars, gas and bulge).  See \cite{bftully} for more details.} for the last data point of the 40 galaxy sample.  The authors chose to express this plot as $v\propto M^{\frac{1}{4}}$ so that the velocities natural error can be retained. The last data point of the galaxy is used since this typically represents the flat region of large spiral galaxy rotation curves, or the end of the rise of dwarf galaxies.  The Tully Fisher plot shows that the data is consistent with the typical $v^4\propto M$ relation (shown in green in the figure), but does show deviation in some of the larger spirals.  This deviation could be explained either by true prediction, or by using an adjusted mass for a given rotation curve. The conformal gravity prediction can be obtained by raising eq. (\ref{total}) to the fourth power while ignoring the quadratic term. One can then take a distance outside the rise and show that $v^4=B(M/M_{\odot})(1+N^*/D))$ with $N^*$ being the dimensionless number of stars in the galaxy.  The constants $B=2c^2M_{\odot}G\gamma_0=0.0074$ km$^4$s$^{-4}$ and $D=\gamma_0/\gamma^*=5.65\times10^{ 10}$ thus showing that for a small dwarf galaxy, $v^4\propto M$ and the deviation comes when the number of stars is competitive with $D$.  Figure \ref{tfplot} shows the full CG prediction in orange.  The fact that conformal gravity predicts a divergence from a pure $v^4\propto M$ is quite important for testing the theory against other possible alternatives.  Since the departure from $v^4$ behavior lies in the largest of spiral galaxies, more recent data such as that found in \cite{highz} could be used to evaluate the limits of the prediction.  This point is doubly important when performing rotation curve physics, since all fitting of rotation curves is sensitive to the distance measurement of the particular galaxy.  Since the Tully Fisher relation has long been established, it is sometimes (usually for dwarf galaxies) assumed and then used to estimate the distance.  This can become a circular argument that could be misleading when testing a theory.  Work on elaborating the departures from Tully Fisher as well as avoiding its use for both mass fitting and distance estimates will be featured in a future work.
\section{Conclusions}
The Radial Acceleration Relation has added a new observation about rotation curve fitting.  It is an empirical feature of the data which can be used to help constrain the dark matter parameter space, as well as challenge an alternative theory which would explain the current dark matter paradigm with new physics.  In this work, we have selected a random, diverse population of galaxies and applied conformal gravity to the set, and have shown that CG can capture the Radial Acceleration Relation.  Further, if we posit that CG is the new physics responsible for the RAR, then the predictions shown in Fig. \ref{cgplots} highlights that CG explains the RAR without the need for dark matter or a fitting function as in eq. (\ref{E1}).  Application of CG to the Baryonic Tully Fisher relation, further establishes how the theory can accommodate all of the current empirical findings in spiral galaxy physics.  Conformal gravity fitting has recently been extended to include the entire SPARC set so that the data can be tested on a one to one basis and will be published in a future paper. The authors would like to thank IWARA for a warm welcome, and to thank the audience at IWARA 2016 who helped inspire this work.  The authors hope that this work highlights the importance of these international meetings and collaborations.

\bibliography{citations2}%

\begin{thebibliography}{}

\bibitem [\protect \citeauthoryear {%
Bettonvil%
, Sutterlin%
, Hammerschlag%
, Rutten%
\BCBL {}\ \BBA {} Stix%
}{%
Bettonvil%
\ \protect \BOthers {.}}{%
{\protect \APACyear {2003}}%
}]{%
Bettonvil2003}
\APACinsertmetastar {%
Bettonvil2003}%
\begin{APACrefauthors}%
Bettonvil, F\BPBI C.%
, Sutterlin, P.%
, Hammerschlag, R\BPBI H.%
, Rutten, R\BPBI J.%
\BCBL {}\ \BBA {} Stix, M.%
\end{APACrefauthors}%
\unskip\
\newblock
\APACrefYearMonthDay{2003}{}{},
\newblock
{\BBOQ}\APACrefatitle {Proc. SPIE Conf. Ser.} {Proc. SPIE Conf. Ser.}{\BBCQ}
\newblock
\BIn{} \BVOL\ 4853, \BPG~306.
\PrintBackRefs{\CurrentBib}

\bibitem [\protect \citeauthoryear {%
Bland-Hawthorn%
, van Breugel%
\BCBL {}\ \BBA {} Gillingham%
}{%
Bland-Hawthorn%
\ \protect \BOthers {.}}{%
{\protect \APACyear {2001}}%
}]{%
Bland2001}
\APACinsertmetastar {%
Bland2001}%
\begin{APACrefauthors}%
Bland-Hawthorn, J.%
, van Breugel, W.%
\BCBL {}\ \BBA {} Gillingham, P\BPBI R.%
\end{APACrefauthors}%
\unskip\
\newblock
\APACrefYearMonthDay{2001}{}{},
\newblock
\unskip
\newblock
\APACjournalVolNumPages{ApJ}{563}{}{611}.
\PrintBackRefs{\CurrentBib}

\bibitem [\protect \citeauthoryear {%
Kosugi%
\ \BBA {} Gillingham%
}{%
Kosugi%
\ \BBA {} Gillingham%
}{%
{\protect \APACyear {2007}}%
}]{%
Kosugi2007}
\APACinsertmetastar {%
Kosugi2007}%
\begin{APACrefauthors}%
Kosugi, T.%
\BCBT {}\ \BBA {} Gillingham, R\BPBI H.%
\end{APACrefauthors}%
\unskip\
\newblock
\APACrefYearMonthDay{2007}{}{},
\newblock
\unskip
\newblock
\APACjournalVolNumPages{Sol. Phys.}{243}{}{3}.
\PrintBackRefs{\CurrentBib}

\bibitem [\protect \citeauthoryear {%
Kosugi%
\ \protect \BOthers {.}}{%
Kosugi%
\ \protect \BOthers {.}}{%
{\protect \APACyear {2009}}%
}]{%
Kosugi2009}
\APACinsertmetastar {%
Kosugi2009}%
\begin{APACrefauthors}%
Kosugi, T.%
, Matsuzaki, K.%
, Sakao, R.%
, Bettonvil, F\BPBI C.%
, Sutterlin, P.%
\BCBL {}\ \BBA {} Hammerschlag, R\BPBI H.%
\end{APACrefauthors}%
\unskip\
\newblock
\APACrefYearMonthDay{2009}{}{},
\newblock
\unskip
\newblock
\APACjournalVolNumPages{Sol. Phys.}{243}{}{3}.
\PrintBackRefs{\CurrentBib}

\bibitem [\protect \citeauthoryear {%
Power%
\ \protect \BOthers {.}}{%
Power%
\ \protect \BOthers {.}}{%
{\protect \APACyear {1975}}%
}]{%
Paivio1975}
\APACinsertmetastar {%
Paivio1975}%
\begin{APACrefauthors}%
Power, J\BPBI D.%
, Cohen, A\BPBI L.%
, Nelson, S\BPBI M.%
\ et al.\end{APACrefauthors}%
\unskip\
\newblock
\APACrefYearMonthDay{1975}{}{},
\newblock
\unskip
\newblock
\APACjournalVolNumPages{Cognition}{37}{2}{635}.
\PrintBackRefs{\CurrentBib}

\bibitem [\protect \citeauthoryear {%
Rutten%
}{%
Rutten%
}{%
{\protect \APACyear {2007}}%
}]{%
Rutten2007}
\APACinsertmetastar {%
Rutten2007}%
\begin{APACrefauthors}%
Rutten, R\BPBI J.%
\end{APACrefauthors}%
\unskip\
\newblock
\APACrefYearMonthDay{2007}{}{},
\newblock
{\BBOQ}\APACrefatitle {The Physics of Chromospheric Plasmas} {The Physics of
  Chromospheric Plasmas}.{\BBCQ}
\newblock
\BIn{} P.~Heinzel, I.~Dorotovic\BCBL {}\ \BBA {} R\BPBI J.~Rutten\ (\BEDS),
  \APACrefbtitle {ASP Conf. Ser.} {ASP Conf. Ser.}\ \BVOL~368, \BPG~27.
\PrintBackRefs{\CurrentBib}

\bibitem [\protect \citeauthoryear {%
Stix%
}{%
Stix%
}{%
{\protect \APACyear {2004}}%
}]{%
Stix2004}
\APACinsertmetastar {%
Stix2004}%
\begin{APACrefauthors}%
Stix, M.%
\end{APACrefauthors}%
\unskip\
\newblock
\APACrefYear{2004},
\newblock
\APACrefbtitle {Astronomy and Astrophysics Library} {Astronomy and Astrophysics
  Library}\ (\PrintOrdinal{2}\ \BEd).
\newblock
\APACaddressPublisher{Berlin}{Springer}.
\PrintBackRefs{\CurrentBib}

\bibitem [\protect \citeauthoryear {%
{Strunk Jr.}%
\ \BBA {} White%
}{%
{Strunk Jr.}%
\ \BBA {} White%
}{%
{\protect \APACyear {1979}}%
}]{%
Strunk1979}
\APACinsertmetastar {%
Strunk1979}%
\begin{APACrefauthors}%
{Strunk Jr.}, W.%
\BCBT {}\ \BBA {} White, E\BPBI B.%
\end{APACrefauthors}%
\unskip\
\newblock
\APACrefYear{1979},
\newblock
\APACrefbtitle {The Elements of Style} {The Elements of Style}\
  (\PrintOrdinal{3}\ \BEd).
\newblock
\APACaddressPublisher{New York}{MacMillan}.
\PrintBackRefs{\CurrentBib}

\end{thebibliography}


\begin{thebibliography}{}

\bibitem [\protect \citeauthoryear {%
{Freeman}%
}{%
{Freeman}%
}{%
{\protect \APACyear {1970}}%
}]{%
freeman}
\APACinsertmetastar {%
freeman}%
\begin{APACrefauthors}%
{Freeman}, K\BPBI C.%
\end{APACrefauthors}%
\unskip\
\newblock
\APACrefYearMonthDay{1970}{{\APACmonth{06}}}{},
\newblock
\unskip
\newblock
\APACjournalVolNumPages{ApJ}{160}{}{811}.
\PrintBackRefs{\CurrentBib}

\bibitem [\protect \citeauthoryear {%
{Genzel}%
\ \protect \BOthers {.}}{%
{Genzel}%
\ \protect \BOthers {.}}{%
{\protect \APACyear {2017}}%
}]{%
highz}
\APACinsertmetastar {%
highz}%
\begin{APACrefauthors}%
{Genzel}, R.%
, {Schreiber}, N\BPBI M\BPBI F.%
, {{\"U}bler}, H.%
\ et al.\end{APACrefauthors}%
\unskip\
\newblock
\APACrefYearMonthDay{2017}{}{},
\newblock
\unskip
\newblock
\APACjournalVolNumPages{\nat}{543}{}{397}.
\PrintBackRefs{\CurrentBib}

\bibitem [\protect \citeauthoryear {%
Lelli%
, McGaugh%
\BCBL {}\ \BBA {} Schombert%
}{%
Lelli%
\ \protect \BOthers {.}}{%
{\protect \APACyear {2016}}%
}]{%
sparc}
\APACinsertmetastar {%
sparc}%
\begin{APACrefauthors}%
Lelli, F.%
, McGaugh, S\BPBI S.%
\BCBL {}\ \BBA {} Schombert, J\BPBI M.%
\end{APACrefauthors}%
\unskip\
\newblock
\APACrefYearMonthDay{2016}{}{},
\newblock
\unskip
\newblock
\APACjournalVolNumPages{The Astronomical Journal}{152}{6}{157}.
\PrintBackRefs{\CurrentBib}

\bibitem [\protect \citeauthoryear {%
{Lelli}%
, {McGaugh}%
, {Schombert}%
\BCBL {}\ \BBA {} {Pawlowski}%
}{%
{Lelli}%
\ \protect \BOthers {.}}{%
{\protect \APACyear {2017}}%
}]{%
onelaw}
\APACinsertmetastar {%
onelaw}%
\begin{APACrefauthors}%
{Lelli}, F.%
, {McGaugh}, S\BPBI S.%
, {Schombert}, J\BPBI M.%
\BCBL {}\ \BBA {} {Pawlowski}, M\BPBI S.%
\end{APACrefauthors}%
\unskip\
\newblock
\APACrefYearMonthDay{2017}{{\APACmonth{02}}}{},
\newblock
\unskip
\newblock
\APACjournalVolNumPages{ApJ}{836}{}{152}.
\PrintBackRefs{\CurrentBib}

\bibitem [\protect \citeauthoryear {%
Mannheim%
\ \BBA {} O'Brien%
}{%
Mannheim%
\ \BBA {} O'Brien%
}{%
{\protect \APACyear {2011}}%
}]{%
impact}
\APACinsertmetastar {%
impact}%
\begin{APACrefauthors}%
Mannheim, P\BPBI D.%
\BCBT {}\ \BBA {} O'Brien, J\BPBI G.%
\end{APACrefauthors}%
\unskip\
\newblock
\APACrefYearMonthDay{2011}{Mar}{},
\newblock
\unskip
\newblock
\APACjournalVolNumPages{Phys. Rev. Lett.}{106}{}{121101}.
\PrintBackRefs{\CurrentBib}

\bibitem [\protect \citeauthoryear {%
{Mannheim}%
\ \BBA {} {O'Brien}%
}{%
{Mannheim}%
\ \BBA {} {O'Brien}%
}{%
{\protect \APACyear {2012}}%
}]{%
fitting}
\APACinsertmetastar {%
fitting}%
\begin{APACrefauthors}%
{Mannheim}, P\BPBI D.%
\BCBT {}\ \BBA {} {O'Brien}, J\BPBI G.%
\end{APACrefauthors}%
\unskip\
\newblock
\APACrefYearMonthDay{2012}{Jun}{},
\newblock
\unskip
\newblock
\APACjournalVolNumPages{Physical Review D}{85}{}{124020}.
\PrintBackRefs{\CurrentBib}

\bibitem [\protect \citeauthoryear {%
{McGaugh}%
}{%
{McGaugh}%
}{%
{\protect \APACyear {2012}}%
}]{%
bftully}
\APACinsertmetastar {%
bftully}%
\begin{APACrefauthors}%
{McGaugh}, S\BPBI S.%
\end{APACrefauthors}%
\unskip\
\newblock
\APACrefYearMonthDay{2012}{{\APACmonth{02}}}{},
\newblock
\unskip
\newblock
\APACjournalVolNumPages{\aj}{143}{}{40}.
\PrintBackRefs{\CurrentBib}

\bibitem [\protect \citeauthoryear {%
McGaugh%
, Lelli%
\BCBL {}\ \BBA {} Schombert%
}{%
McGaugh%
\ \protect \BOthers {.}}{%
{\protect \APACyear {2016}}%
}]{%
mcgaughprl}
\APACinsertmetastar {%
mcgaughprl}%
\begin{APACrefauthors}%
McGaugh, S\BPBI S.%
, Lelli, F.%
\BCBL {}\ \BBA {} Schombert, J\BPBI M.%
\end{APACrefauthors}%
\unskip\
\newblock
\APACrefYearMonthDay{2016}{Nov}{},
\newblock
\unskip
\newblock
\APACjournalVolNumPages{Phys. Rev. Lett.}{117}{}{201101}.
\PrintBackRefs{\CurrentBib}

\bibitem [\protect \citeauthoryear {%
{Navarro}%
, {Frenk}%
\BCBL {}\ \BBA {} {White}%
}{%
{Navarro}%
\ \protect \BOthers {.}}{%
{\protect \APACyear {1996}}%
}]{%
nfw}
\APACinsertmetastar {%
nfw}%
\begin{APACrefauthors}%
{Navarro}, J\BPBI F.%
, {Frenk}, C\BPBI S.%
\BCBL {}\ \BBA {} {White}, S\BPBI D\BPBI M.%
\end{APACrefauthors}%
\unskip\
\newblock
\APACrefYearMonthDay{1996}{{\APACmonth{05}}}{},
\newblock
\unskip
\newblock
\APACjournalVolNumPages{\apj}{462}{}{563}.
\PrintBackRefs{\CurrentBib}

\bibitem [\protect \citeauthoryear {%
O'Brien%
, Chiarelli%
\BCBL {}\ \BBA {} Mannheim%
}{%
O'Brien%
\ \protect \BOthers {.}}{%
{\protect \APACyear {2018}}%
}]{%
jgoprb}
\APACinsertmetastar {%
jgoprb}%
\begin{APACrefauthors}%
O'Brien, J.%
, Chiarelli, T.%
\BCBL {}\ \BBA {} Mannheim, P.%
\end{APACrefauthors}%
\unskip\
\newblock
\APACrefYearMonthDay{2018}{05}{},
\newblock
\unskip
\newblock
\APACjournalVolNumPages{Physics Letters B}{782}{}{}.
\PrintBackRefs{\CurrentBib}

\end{thebibliography}

\end{document}